\documentclass[12pt]{iopart} 
\pdfoutput=1

\usepackage{iopams}

\usepackage[utf8]{inputenc}
\usepackage{lmodern}
\expandafter\let\csname equation*\endcsname\relax

\expandafter\let\csname endequation*\endcsname\relax
\usepackage{amsmath}
\usepackage{graphicx}
\usepackage{float}
\usepackage{amsfonts}
\usepackage{url}
\usepackage{epstopdf}
\usepackage{placeins}
\usepackage{xcolor}
\usepackage{verbatim}
\usepackage{placeins}
\usepackage{cite}

\usepackage[cal=cm]{mathalfa}

\usepackage{color}   
\usepackage{hyperref}
\hypersetup{
    colorlinks=true, 
    linktoc=all,     
    linkcolor=red,  
}

\begin{document}

\title[Dynamical Generation of Synthetic Electric Fields in the Quantum Regime]{Dynamical Generation of Synthetic Electric Fields for Photons  in the Quantum Regime}

\author{Petr Zapletal and Andreas Nunnenkamp}
\address{Cavendish Laboratory, University of Cambridge, Cambridge CB3 0HE, United Kingdom}

 \eads{\mailto{pz270@cam.ac.uk}}

\begin{abstract}
Optomechanics offers a natural way to implement synthetic dynamical gauge fields, leading to synthetic electric fields for phonons and, as a consequence, to unidirectional light transport.
Here we investigate the quantum dynamics of synthetic gauge fields in the minimal setup of two optical modes coupled by phonon-assisted tunneling where the phonon mode is undergoing self-oscillations.
We use the quantum van-der-Pol oscillator as the simplest dynamical model for a mechanical self-oscillator that allows us to perform quantum master equation simulations. We identify a single parameter, which controls the strength of quantum fluctuations, enabling us to investigate the classical-to-quantum crossover. 
We show that the generation of synthetic electric fields is robust against noise and that it leads to unidirectional transport of photons also in the quantum regime, albeit with a reduced isolation ratio.
Our study opens the path for studying dynamical gauge fields in the quantum regime based on optomechanical arrays.
\end{abstract}


\maketitle

\section{Introduction}
Optomechanics describes the interaction of light and matter due to radiation pressure \cite{aspelmeyer2014}. The field of optomechanics has produced many experimental achievements in the past decade such as cooling a mechanical oscillator into its motional ground state \cite{chan2011,teufel2011}, position measurement with a precision below that at the standard quantum limit \cite{wilson2015}, coherent state transfer \cite{palomaki2013a}, squeezing of a mechanical oscillator \cite{wollman2015,pirkkalainen2015,lecocq2015}, manipulation of single phonons \cite{hong2017} and entanglement of mechanical oscillators \cite{riedinger2018,ockeloen-korppi2018}.

Implementing synthetic gauge fields for neutral particles has recently  attracted a lot of attention \cite{goldman2014,hartmann2016}.
They have been proposed and realized using cold atoms \cite{jaksch2003,lin2009,aidelsburger2011,aidelsburger2013}, photons \cite{haldane2008,wang2009,hafezi2011,fang2012,rechtsman2013,hafezi2013,mittal2014} and phonons \cite{peano2015,nash2015,susstrunk2015,wang2015a,brendel2017,brendel2018,seif2018}. It has been shown that synthetic gauge fields can naturally arise in optomechanical setups \cite{hafezi2012,schmidt2015}: due to the optomechanical interaction, photons pick up the phase of a mechanical oscillator and the mechanical phase then represents a synthetic gauge field for photons with $U(1)$ gauge symmetry. The generation of artificial magnetic fields has led, together with reservoir engineering \cite{metelmann2015}, to the recent groundbreaking experimental work on optomechanical nonreciprocity \cite{kim2015,wang2015,ruesink2016,fang2017,bernier2017,barzanjeh2017}. For future realizations of optomechanical arrays \cite{chang2011,heinrich2011,xuereb2012,ludwig2013}, synthetic gauge fields have been proposed to give rise to topological phases of sound and light \cite{schmidt2015} including topologically-protected phonon transport \cite{peano2015}.

Most of the above realizations lead to \emph{static} synthetic gauge fields, which are fixed externally by the driving. The possibility to implement \emph{synthetic dynamical gauge fields}, i.e.~fields that are dynamical degrees of freedom themselves, has been considered in ultra-cold atoms in optical lattices \cite{banerjee2012,zohar2016}, superconducting circuits \cite{marcos2013,marcos2014}, cavity quantum electrodynamics \cite{ballantine2017}, and trapped ions \cite{hauke2013,martinez2016}. It has been shown very recently that the optomechanical interaction offers a natural way to engineer synthetic dynamical gauge fields \cite{walter2016}. In that case, a mechanical oscillator is driven into self-sustained oscillations. In this way, the phase of the mechanical oscillator evolves according to its own dynamics and can provide a dynamical gauge field for photons. The dynamics of synthetic gauge fields in open one-dimensional arrays has been shown to give rise to synthetic electric fields for photons \cite{zapletal2018}. The dynamical generation of these synthetic electric fields depends on the direction of light propagation. This gives rise to unidirectional light transport: light propagation is exponentially suppressed in one direction whereas light can propagate freely in the reverse direction.


In this manuscript, we study the minimal setup exhibiting synthetic dynamical gauge fields: two optical modes that are coupled by phonon-assisted tunneling where the phonon mode is undergoing self-oscillations. Cavity optomechanics is approaching the nonlinear quantum regime where a single photon displaces the mechanical oscillator by more than its zero-point uncertainty \cite{rabl2011,nunnenkamp2011}. This has motivated the theoretical investigation of optomechanically induced self-oscillations in the quantum regime, showing that self-oscillations persist even if only a few energy levels are occupied \cite{ludwig2008, vahala2008, rodrigues2010, armour2012, qian2012, lorch2014}. In particular, we use the so-called quantum van-der-Pol oscillator \cite{lee2013,walter2014} as a model of mechanical self-oscillations. We show that in our model the strength of quantum fluctuations is controlled by a single parameter. By tuning the strength of quantum fluctuations, we explore the classical-to-quantum crossover, an analysis which has been carried out for the standard optomechanical setup \cite{ludwig2008,weiss2016}. We show that the phase of the mechanical oscillator is very sensitive to noise and as a consequence its value is washed out deep in the quantum regime by quantum fluctuations. However, the mechanical phase still represents a synthetic gauge field for photons, despite being very noisy, since quantum fluctuations do not break the gauge symmetry of the system. We show that even though the value of the mechanical phase is washed out in the quantum regime, its nonlinear dynamics still results in the generation of synthetic electric fields. This demonstrates the robustness of synthetic electric fields against noise, which emerge via a dynamical instability. In contrast to the classical regime, the transition from a vanishing to a finite synthetic electric field is blurred by quantum fluctuations \cite{drummond1980}. Sufficiently above threshold, the synthetic electric field is not affected by fluctuations and thus also the resulting unidirectional light transport is robust against noise. However, the quantization of the van-der-Pol oscillator's energy levels leads to an increased mechanical amplitude with respect to the classical regime, and an increased transmission in the blockaded direction. This results in reduced isolation ratios, which, however, do not preclude unidirectional light transport in arrays.

Our work shows that cavity optomechanics provides a suitable platform for studying the quantum dynamics of synthetic gauge fields. In the future, it will allow for the investigation of genuine quantum features in synthetic gauge fields and the interplay of their dynamics with quantum synchronization \cite{lee2013,walter2014,weiss2016,lorch2017}.

\section{Model}

\begin{figure}[t]
\centering \includegraphics[width=0.444\linewidth]{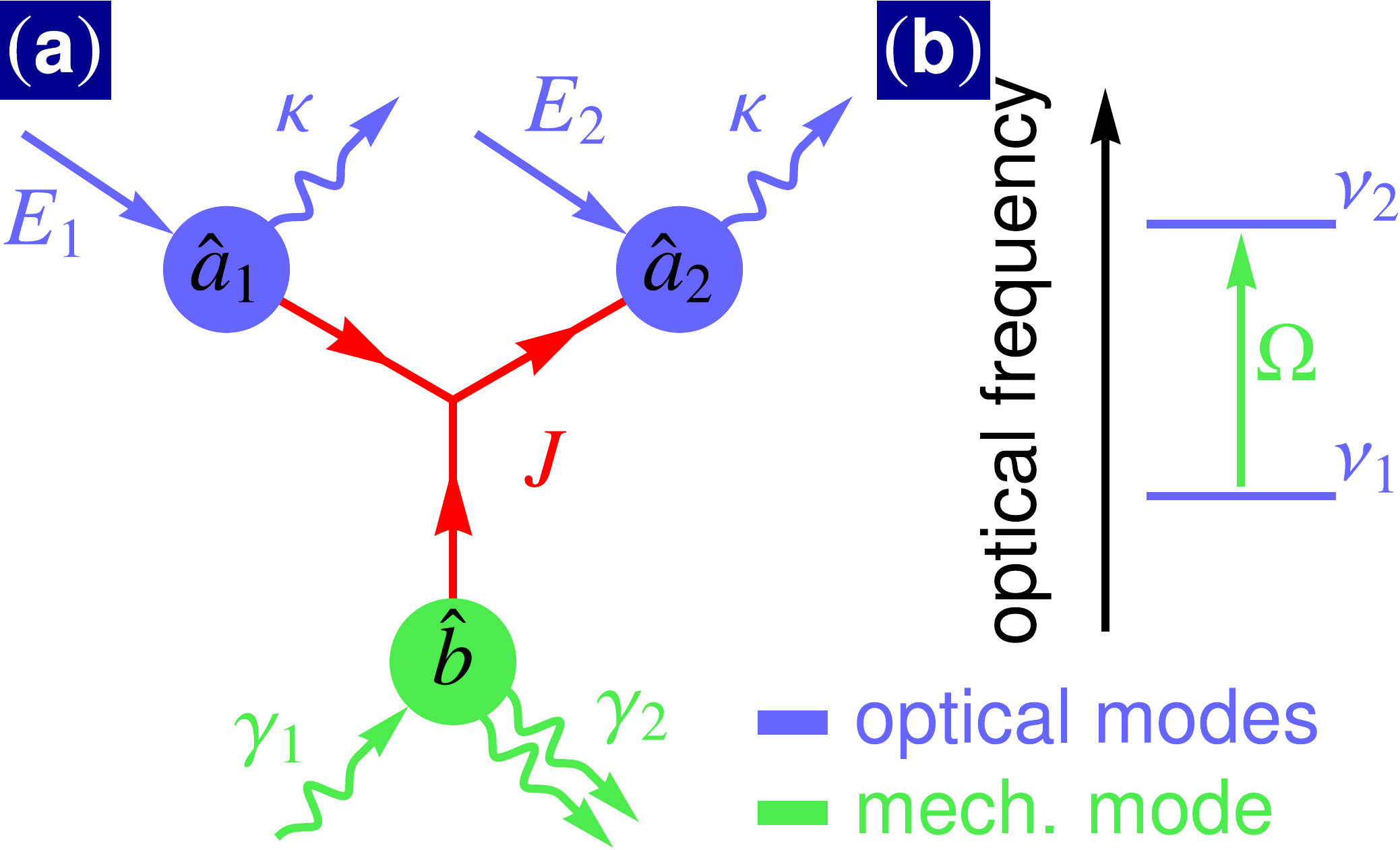} \caption{Schematic of an optomechanical setup exhibiting a synthetic dynamical gauge field for photons. (a) Red arrows depict tunneling of photons (with tunneling amplitude $J$) from the optical mode, $\hat{a}_1$, to the optical mode, $\hat{a}_2$, assisted by the coherent absorption of a phonon from the mechanical mode $\hat{b}$. The inverse process happens with the same tunneling amplitude $J$. The mechanical oscillator performs self-oscillations arising from the competition of linear (one-phonon) anti-damping and nonlinear (two-phonon) damping with rates $\gamma_1$ and $\gamma_2$, respectively. Photons decay at a rate $\kappa$. Optical modes are driven with lasers of amplitudes $E_1$ and $E_2$ to study the transmission through the device. (b) Optical frequencies $\nu_1$ and $\nu_2$ satisfy the resonance condition $\Omega\approx\nu_2-\nu_1$, where $\Omega$  is the mechanical frequency.}
\label{link} 
\end{figure}

Dynamical gauge fields for photons have been proposed for optomechanical arrays with phonon-assisted photon tunneling \cite{walter2016}. The minimal setup of this kind, as shown in figure~\ref{link}, consists of two optical modes and one mechanical mode. 
The Hamiltonian describing this setup (with $\hbar=1$) is
\begin{equation}\label{ham}
\hat{H}= -\sum_{j=1}^2\Delta_{O,j} \hat{a}^{\dagger}_{j} \hat{a}_{j} - \Delta_M \,\hat{b}^{\dagger} \hat{b} + J \left(\hat{b}\hat{a}^{\dagger}_{2} \hat{a}_{1} + {\rm h.c.} \right) + \sum_{j=1}^2E_j(\hat{a}_j+\hat{a}_j^{\dagger}),
\end{equation}
where $\Delta_{O,j}=\nu_j-\nu_{D,j}$ and $\Delta_M=\Omega-\nu_{D,2}+\nu_{D,1}$ are optical and mechanical detunings \cite{walter2016}. In the third term, $\hat{b}\hat{a}^{\dagger}_{2} \hat{a}_{1}$, with amplitude $J$, a mechanical mode $\hat{b}$ with frequency $\Omega$ assists the tunneling between two optical modes $\hat{a}_1$ and $\hat{a}_2$ with frequencies $\nu_1$ and $\nu_2$, respectively. This process can be spectrally selected by satisfying the resonance condition, $\Omega\approx\nu_2-\nu_1$. The optical modes $\hat{a}_j$ are coherently driven by lasers at frequencies $\nu_{D,j}$ and with strengths $E_j$. 
The optical and mechanical modes are expressed in suitable rotating frames and fast rotating terms are neglected in the Hamiltonian (\ref{ham}) within the rotating-wave approximation, valid for $\kappa, J, J B\ll \Omega$, where $\kappa$ is the photon decay rate and $B=|\langle \hat{b}\rangle|$ is the mechanical amplitude.

To implement dynamical gauge fields, the mechanical oscillator is assumed, as in \cite{walter2016}, to perform self-sustained oscillations. In this case, the phase $\phi$ of the mechanical oscillator is able to evolve according to its own dynamics which provides a dynamical gauge field for photons with a $U(1)$ symmetry. Any time evolution of the mechanical phase can be seen as generating a synthetic electric field $\mathcal{E}=\dot{\phi}$ for the photons \cite{zapletal2018} (see \ref{el field} for more details).

In this manuscript, we employ the quantum van-der-Pol oscillator as a model of the mechanical self-oscillations enabling us to study synthetic dynamical gauge fields in the quantum regime. Two optical modes with tunneling assisted by such a mechanical self-oscillator are described by the following quantum master equation in Lindblad form
\begin{equation}\label{master}
\dot{\hat{\rho}}=-i\left[\hat{H},\hat{\rho}\right]+\kappa \sum_{j=1}^2\mathcal{D}\left[\hat{a}_j\right]\hat{\rho} + \gamma_1\mathcal{D}[\hat{b}^{\dagger}]\hat{\rho} + \gamma_2\mathcal{D}[\hat{b}^2]\hat{\rho},
\end{equation}
for the composite density matrix $\rho$, where $\mathcal{D}[\hat{O}]\hat{\rho}=\hat{O}\hat{\rho}\hat{O}^{\dagger}-\frac{1}{2}\left\{\hat{O}^{\dagger}\hat{O},\hat{\rho}\right\}$ and $\left\{\cdot,\cdot\right\}$ is the anti-commutator. The Hamiltonian (\ref{ham}) governs the coherent dynamics of the system.  Photon decay at a rate $\kappa$ is described by the dissipator  $\mathcal{D}[\hat{a}_j]$. Self-oscillations of the mechanical mode arise from the competition of two dissipative processes, namely negative damping, $\mathcal{D}[\hat{b}^{\dagger}]$, and nonlinear damping, $\mathcal{D}[\hat{b}^2]$, with rates $\gamma_1$ and $\gamma_2$, respectively. Mechanical self-oscillations persist even in the extreme quantum limit, $\gamma_2/\gamma_1\rightarrow\infty$, where only two lowest energy levels are occupied \cite{lee2013,walter2014}. As a result, the model described by the master equation (\ref{master}) allows us to study the dynamics of synthetic gauge fields deep in the quantum regime for low occupations of the mechanical self-oscillator. The master equation (\ref{master}) is a starting point of our analysis.


\section{Classical limit and quantum parameter}
We start our analysis by considering the classical regime of large mechanical and optical occupations  when quantum fluctuations are negligible.

In the classical regime, quantum fluctuations and the quantization of energy levels can be neglected for the van-der-Pol oscillator \cite{walter2014} as well as for the optical modes. The system can then be described by equations of motion for the expectation values $\langle \hat{b} \rangle={\rm Tr}\left[\hat{\rho}\hat{b} \right]\equiv b$, $\langle \hat{a}_1 \rangle={\rm Tr}\left[\hat{\rho}\hat{a}_1 \right] \equiv a_1$, $\langle \hat{a}_2 \rangle={\rm Tr}\left[\hat{\rho}\hat{a}_2 \right] \equiv a_2$, derived from the quantum master equation (\ref{master}).
In the classical regime, it is a good approximation to factorize the higher order moments, ${\rm Tr}\left[\hat{\rho}\hat{b}^{\dagger}\hat{b}^2 \right]\approx |b|^2 b$ and ${\rm Tr}\left[\hat{\rho}\hat{a}_1^{\dagger}\hat{a}_2 \right]\approx a_1^* a_2$. In this way we obtain a set of classical equations of motion
\begin{align}
\dot{b} & =i \Delta_M b - iJ a^*_1 a_2 +\frac{\gamma_1}{2}b-\gamma_2|b|^2b,\label{EOM b}\\
\dot{a}_k & = i \Delta_{\rm O} a_k - i E - i J \left(\delta_{1,k}b^*+\delta_{2,k}b \right) a_{l} -\frac{\kappa}{2}a_k,\\
\dot{a}_l & = - i J \left(\delta_{1,k}b+\delta_{2,k}b^*\right)a_{k} -\frac{\kappa}{2}a_l\label{a2 tilde},
\end{align}
where we assume that only one optical mode, $a_k$, is driven,  $\Delta_{O} = \Delta_{O,k}$ and $\delta_{j,k}$ is the Kronecker delta. The detuning, $\Delta_{O,l}$, of the non-driven optical mode, $a_l$, ($l\neq k$) can be set to zero in the absence of the laser drive. Using rescaled optical amplitudes $\tilde{a}_{j}=\kappa a_j/E$, mechanical amplitude $\tilde{b}=J b / \kappa$ and time  $\tilde{t}=\kappa t$, the equations of motion can be expressed as \begin{align}
\frac{{\rm d}\tilde{b}}{{\rm d}\tilde{t}} & =i\tilde{\Delta}_M \tilde{b} - i \mathcal{P} \tilde{a}^*_1 \tilde{a}_2 +\frac{\tilde{\gamma}_1}{2}\tilde{b}-\tilde{\gamma}_2|\tilde{b}|^2\tilde{b},\label{eq_bt}\\
\frac{{\rm d}\tilde{a}_k}{{\rm d}\tilde{t}} & = i \tilde{\Delta}_{\rm O} \tilde{a}_k - i - i \left(\delta_{1,k}\tilde{b}^*+\delta_{2,k}\tilde{b} \right) \tilde{a}_{l} -\frac{1}{2}\tilde{a}_k,\label{eq_akt}\\
\frac{{\rm d}\tilde{a}_l}{{\rm d}\tilde{t}} & = - i \left(\delta_{1,k}\tilde{b}+\delta_{2,k}\tilde{b}^*\right)\tilde{a}_{k} -\frac{1}{2}\tilde{a}_l,\label{eq_alt}
\end{align}
which depend on the dimensionless parameters $\tilde{\Delta}_O=\Delta_O/\kappa$, $\tilde{\Delta}_M=\Delta_M/\kappa$, $\tilde{\gamma}_1=\gamma_1/\kappa$, $\tilde{\gamma_2}=\kappa\gamma_2/J^2$ as well as the rescaled driving power $\mathcal{P}=J^2E^2/\kappa^4$. 

However, the classical equations of motion do not depend on the quantum parameter
\begin{equation}
\zeta=\frac{J}{\kappa},
\end{equation}
which has been shown to control the strength of quantum fluctuations in the standard optomechanical setup \cite{ludwig2008,weiss2016}. Since $\zeta$ determines the scaling of the mechanical amplitude, it controls the strength of quantum fluctuations for the van-der-Pol oscillator. The quantum parameter $\gamma_2/\gamma_1$ of the van-der-Pol oscillator  \cite{walter2014} can be re-expressed as
\begin{equation}
\frac{\gamma_2}{\gamma_1}=\frac{\tilde{\gamma}_2}{\tilde{\gamma}_1}\zeta^2
\end{equation}
in terms of the standard optomechanical quantum parameter $\zeta$ and the two classical parameters $\tilde{\gamma}_1$ and $\tilde{\gamma}_2$. As a result, the strength of quantum fluctuations in our model with the quantum van-der-Pol oscillator is controlled by the quantum parameter $\zeta$. 

In the classical regime, where quantum fluctuations are negligible, the optical and mechanical amplitudes can be rescaled and the system dynamics is described by the rescaled equations of motion (\ref{eq_bt}), (\ref{eq_akt}) and (\ref{eq_alt}). These classical equations are similar to those studied in \cite{zapletal2018} giving rise to the synthetic electric field for photons
\begin{equation}
\tilde{\mathcal{E}}=\frac{{\rm d}\phi}{{\rm d}\tilde{t}}=\tilde{\Delta}_M - \mathcal{P}\frac{|\tilde{a}_1||\tilde{a}_2|}{|\tilde{b}|}\cos\left(\phi - \theta_2 + \theta_1\right),\label{el class}
\end{equation}
where $\theta_j$ is the phase of the optical mode $a_j$. Further to the analysis presented in \cite{zapletal2018}, the effects of radiation pressure on the mechanical amplitude are taken into account in (\ref{eq_bt}). This modification of the mechanical amplitude does not qualitatively change the steady state of the model (for a detailed analysis, see \ref{steady vdp}).
If light propagates from the lower to the higher optical frequency, a synthetic electric field is generated above a threshold driving power, and, as a consequence, light transport is suppressed. The synthetic electric field vanishes if light propagates from the higher to the lower optical frequency. This leads to unidirectional light transport via the dynamically generated synthetic electric field studied in \cite{zapletal2018}.


\section{Synthetic electric fields in the quantum regime}
Now we study the effect of quantum fluctuations both close to the classical limit, $\zeta\rightarrow0$, as well as deep in the quantum regime, $\zeta\gtrsim1$. 

We investigate steady states of the system in the resonant case, i.e., $\Delta_M=\Delta_O=0$, for which the effects, studied here, are most pronounced.
In this manuscript, we consider driving the lower optical frequency, where the synthetic electric field is dynamically generated and light transmission is suppressed \cite{zapletal2018}.
In the case of driving the higher optical frequency, no synthetic electric field is generated and light can freely propagate to the lower optical frequency (see \ref{higher freq} for more details).

To understand how the dimensionless parameters introduced in the last section influence the dynamics of the quantum model, we rewrite the master equation (\ref{master}) in terms of these parameters. The master equation then reads
\begin{equation}\label{master_res}
\frac{{\rm d}\hat{\rho}}{{\rm d}\tilde{t}}=-i\left[\hat{\mathcal{H}},\hat{\rho}\right] + \sum_{j=1}^2\mathcal{D}\left[\hat{a}_j\right]\hat{\rho} + \tilde{\gamma}_1\mathcal{D}[\hat{b}^{\dagger}]\hat{\rho} + \zeta^2\tilde{\gamma}_2\mathcal{D}[\hat{b}^2]\hat{\rho},
\end{equation}
where
\begin{equation}\label{ham_res}
\hat{\mathcal{H}}= \zeta \left(\hat{b}\hat{a}^{\dagger}_{2} \hat{a}_{1} + {\rm h.c.} \right) + \frac{\sqrt{\mathcal{P}}}{\zeta}(\hat{a}_1+\hat{a}_1^{\dagger}).
\end{equation}
We can see that the quantum parameter $\zeta$, controls the scaling of optical amplitudes, since the driving strength is $\sqrt{\mathcal{P}}/\zeta$. It also determines the scaling of the mechanical amplitude which is proportional to the ratio of negative damping rate, $\tilde{\gamma}_1$, and the nonlinear damping rate, $\zeta^2\tilde{\gamma}$. As we have shown in the last section, the scaling of optical and mechanical amplitudes does not influence the classical dynamics of the system. The absolute strength of quantum fluctuations, $\left[\hat{b},\hat{b}^{\dagger} \right] = 1$, does not depend on the scaling of the mechanical amplitude, $\sqrt{\langle \hat{b}^{\dagger} \hat{b}\rangle}\propto1/\zeta$. As a result, the relative strength of quantum fluctuations compared to the mechanical amplitude increases with $\zeta$.

Since time is now expressed in the units of photon decay rate $\kappa$, the dimensionless parameter $\tilde{\gamma}_1$ represents the ratio of the optical coherence time and the mechanical coherence time. This parameter determines to which extent the mechanical amplitude is influenced by the coupling to the optical modes. We choose $\tilde{\gamma}_1=10$, in which case the mechanical dissipative processes are fast compared to the dynamics of optical modes and they dominate the dynamics of the system. In this regime, the modification of the mechanical amplitude due to the coupling to optical modes is marginal. On the other hand, the mechanical phase undergoes a slow dynamics which arise from the coupling to the optical modes since the mechanical dissipative processes are phase insensitive.  In this regime, we can study the dynamics of the synthetic gauge field, i.e. the mechanical phase, since the effects originating from the dynamics of mechanical amplitude are eliminated.


We consider the relative phase, $\varphi=\phi - \theta_2 + \theta_1$, between the mechanical mode of phase $\phi$ and the optical modes of phases $\theta_j$ to study the generation of synthetic electric fields. The relative phase $\varphi$ determines the action of the radiation pressure force on the mechanical phase and as a consequence also the value of the synthetic electric field as it can be read off from the classical equation (\ref{el class}). The distribution 
\begin{equation}
P(\varphi)=\iiint_{0}^{2\pi}{\rm d}\phi{\rm d}\theta_1{\rm d}\theta_2\,\delta(\phi - \theta_2 + \theta_1 - \varphi)p(\phi,\theta_1,\theta_2)
\end{equation}
of the relative phase can be calculated from the steady-state density matrix $\hat{\rho}_{SS}$, where
\begin{equation}
p( \phi, \theta_1,\theta_2)= {\rm Tr}\left[|\phi, \theta_1,\theta_2\rangle\langle \phi,\theta_1 \theta_2 |\hat{\rho}_{\rm SS}\right],
\end{equation}
$|\phi, \theta_1,\theta_2\rangle=\frac{1}{(2\pi)^{3/2}}\sum_{n,m,l=0}^{\infty}e^{i\left(\,n\phi+m\theta_1+l \theta_2\right)}|n,m,l\rangle$, $|n\rangle$ is the Fock state. To study the effect of fluctuations on the relative phase $\varphi$, we employ
\begin{equation}
M=2\pi \max\left(P(\varphi)\right)-1,
\end{equation}
as a measure, which was used in \cite{hush2015,lorch2017}. The quantity, $M$, is the height of a peak in the relative phase distribution, $P(\varphi)$. In the following, we will refer to $M$ as the phase coherence measure.


\begin{figure}[t]
\centering
\includegraphics[width=0.9\linewidth]{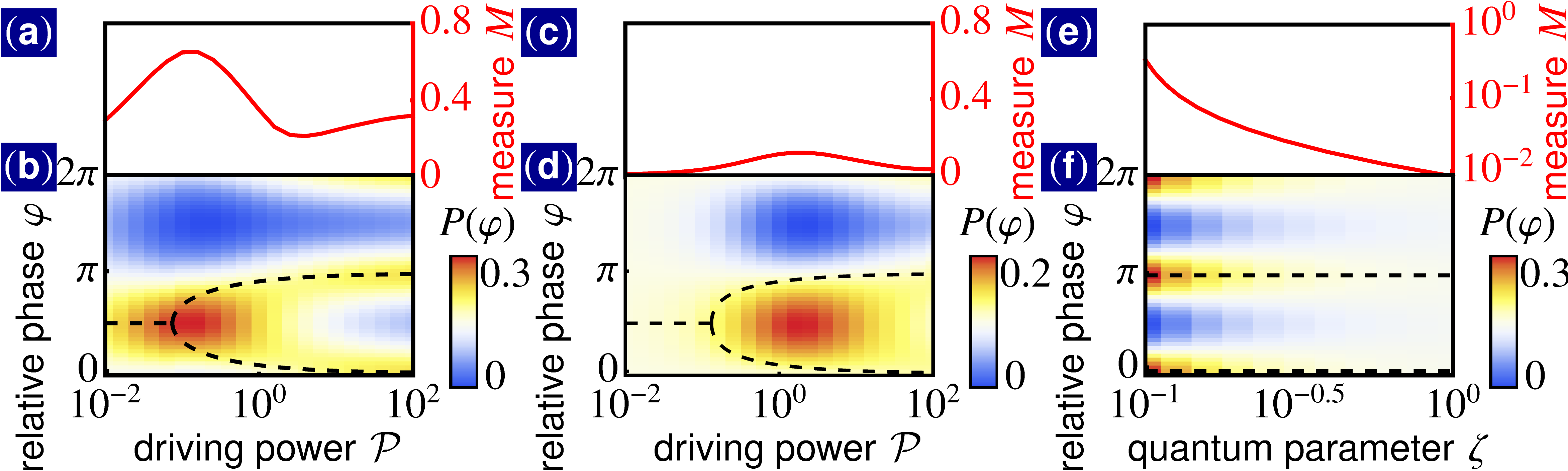}
\caption{Relative phase between mechanical and optical modes if the lower optical frequency is driven. Relative phase distribution as a function of driving power $\mathcal{P}$ at the onset of a finite synthetic electric field: (b) close to the classical limit and (d) in the quantum regime. Corresponding phase coherence measure: (a) close to the classical limit and (c) in the quantum regime. Classical-to-quantum crossover: (f) relative phase distribution (shown on a logarithmic scale) and (e) phase coherence measure as a function of the quantum parameter $\zeta$ for a large synthetic electric field. The black dashed lines show the classical value of the relative phase. (Master equation simulations for: (a) and (b) $\tilde{\gamma}_1=10$, $\tilde{\gamma}_2=200$, and $\zeta = 0.1$; (c) and (d) $\tilde{\gamma}_1=10$, $\tilde{\gamma}_2=40$, and $\zeta = 1$; and (e) and (f) $\tilde{\gamma}_1=10$, $\tilde{\gamma}_2=200$, and $\mathcal{P}=10000$.)}
\label{Pphi}
\end{figure}

In figure~\ref{Pphi}, we show the relative phase distribution and the phase coherence measure as a function of driving power, $\mathcal{P}$, close to the classical limit and deep in the quantum regime. In the classical limit, the relative phase (black dashed lines) has a single value $\varphi=\pi/2$ at small driving power, see figure~\ref{Pphi}(b), indicating that the synthetic electric field vanishes in steady state (see \ref{steady vdp}). For sufficiently large driving powers, the relative phase bifurcates to values different from $\pi/2$ and a finite synthetic electric field  is generated \cite{zapletal2018}. 

While the classical equations of motion neglect fluctuations in the relative phase, the master equation simulation (see \ref{simulations} for more details) takes these fluctuations into account and we study them using the relative phase distribution. Close to the classical limit, the relative phase distribution is peaked around the classical value of the relative phase, see figure~\ref{Pphi}(b). Fluctuations of the relative phase are reduced in the vicinity of the classical threshold value of driving power. This can be seen from the sharper peak in the relative phase distribution around the classical value (figure~\ref{Pphi}(b)) and the increase of the phase coherence measure at the onset of a finite synthetic electric field, see figure~\ref{Pphi}(a). In the classical limit, the steady state is bistable above threshold. Depending on the initial conditions, the system ends up in one of the stationary solutions of the classical equations (\ref{eq_bt}), (\ref{eq_akt}) and (\ref{eq_alt}). Fluctuations lead to transitions between these two classically stable stationary solutions and to a unique steady state distribution of the relative phase, which is peaked around the two classical stationary values.

\begin{figure}[t]
\centering
\includegraphics[width=0.9\linewidth]{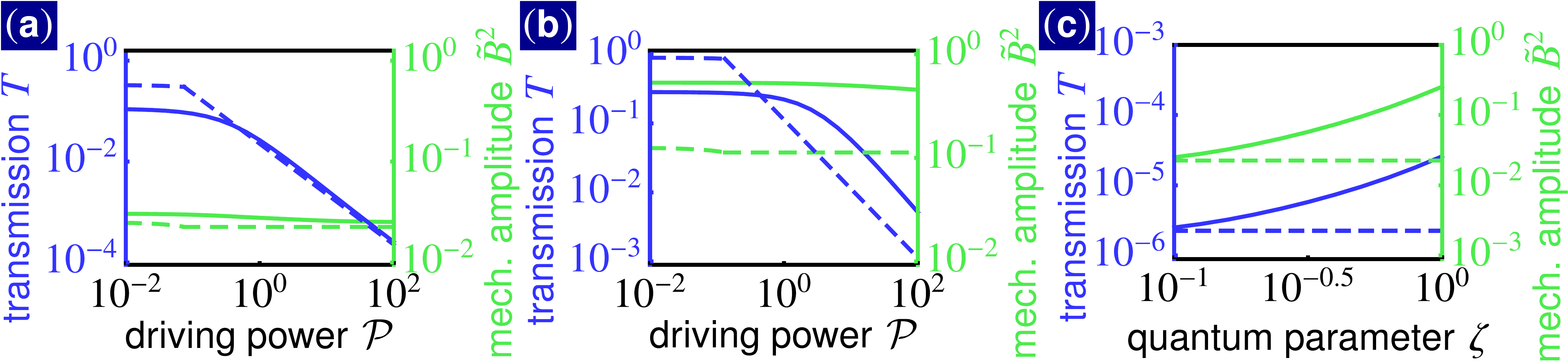}
\caption{Light propagation if the lower optical frequency is driven.
Optical transmission $T$ and expectation value of the mechanical amplitude $\tilde{B}^2=\zeta^2\langle\hat{b}^{\dagger}\hat{b}\rangle$ as a function of driving power $\mathcal{P}$ at the onset of a finite synthetic electric field: (a) close to the classical limit and (b) deep in the quantum regime.
(c) Classical-to-quantum crossover: optical transmission and mechanical amplitude as a function of the quantum parameter for large synthetic electric fields.
The dashed lines show classical limit.
(Master equation simulations for: (a) $\tilde{\gamma}_1=10$, $\tilde{\gamma}_2=200$, and $\zeta = 0.1$; (b) $\tilde{\gamma}_1=10$, $\tilde{\gamma}_2=40$, and $\zeta = 1$; and (c) $\tilde{\gamma}_1=10$, $\tilde{\gamma}_2=200$, and $\mathcal{P}=10000$.)}
\label{trans}
\end{figure}

The dynamics deep in the quantum regime shares several features with that in the classical regime, see figure~\ref{Pphi}(d). At small driving power, the relative phase distribution is peaked around $\varphi=\pi/2$ indicating a vanishing synthetic electric field. For sufficiently large driving powers, the synthetic electric field is generated leading to a bimodal relative phase distribution. The onset of a finite synthetic electric field is shifted to larger driving powers compared to the classical case (black dashed lines). This can be seen both in the relative phase distribution in figure~\ref{Pphi}(d) and the phase coherence measure in figure~\ref{Pphi}(c). The shifted onset of a finite synthetic electric field is due to a special feature of the quantum van-der-Pol oscillator: the nonlinear (two-phonon) damping cannot dissipate the last remaining phonon \cite{walter2014}. This is not captured by the classical equations of motion. As a result, the mechanical amplitude according to the full quantum model (green solid line) is larger than classically expected,  see figure~\ref{trans}(b). Since the mechanical amplitude has to be overcome by the lower-frequency optical amplitude to generate the synthetic electric field \cite{zapletal2018}, the onset of a finite synthetic electric field is shifted to larger driving powers.


If a synthetic electric field is generated, it leads to a suppression of light propagation to the higher-frequency optical mode. In figure~\ref{trans}, optical transmission $T$ is shown as a function of driving power $\mathcal{P}$.
The transmission $T$ is the ratio of the output power leaking from the higher-frequency optical mode, $\kappa\langle\hat{a}_2^{\dagger} \hat{a}_2\rangle$, to the driving power, $\kappa \mathcal{P}/\zeta^2$. Close to the classical limit, the transmission $T$ is slightly decreased for small driving powers. This is due to fluctuations broadening the mechanical frequency by more than the optical linewidth so that the resonant condition, $\Omega = \nu_2 - \nu_1$, is not satisfied for every scattering event (absorption of a phonon). For large driving powers, the generated synthetic electric field significantly suppresses the transmission $T$ by many orders of magnitude.
A similar dependence of the transmission $T$ on driving power is found in the quantum regime, see figure~\ref{trans}(b). For small driving powers, the transmission is marginally reduced by fluctuations. For sufficiently large driving powers, the transmission is strongly suppressed by the dynamically generated synthetic electric field. However, the same amount of suppression occurs for larger driving powers than in the classical limit since the onset of a finite synthetic electric field is shifted due to larger mechanical amplitude $\tilde{B}^2$.

\section{Classical-to-quantum crossover}

To shed more light on the influence of quantum fluctuations and the quantization of energy levels, we study transmission across the classical-to-quantum crossover. 

We consider driving powers well above the onset of a finite synthetic electric field. The synthetic electric field then remains constant as the quantum regime is approached for increasing $\zeta$. As a result, the relative phase distribution $P(\varphi)$ is always peaked around its classical value, see figure~\ref{Pphi}(f). On the other hand, as the quantum regime is reached, the relative phase distribution is washed out by quantum fluctuations. As a consequence, the phase coherence measure significantly decreases, as seen in figure~\ref{Pphi}(e). 

This demonstrates the robustness of the synthetic electric field generation against noise. According to the classical equations of motion, the synthetic electric field, $\tilde{\mathcal{E}}\propto - \cos\left(\varphi\right)$, depends on the relative phase, $\varphi$, see (\ref{el class}). 
The relative phase distribution is washed out by quantum fluctuations as the quantum regime is approached. However, the dynamics of the mechanical phase still results in the generation of the synthetic electric field. It causes a shift in the frequency of the non-driven optical mode, which can be seen in the power spectrum of this mode (shown in \ref{app spectrum}). As a result, the dynamically generated synthetic electric electric field detunes the phonon-assisted photon tunneling from resonance leading to a suppression of light transmission even in the quantum regime (see figure~\ref{trans}(c)).
This shows that while the values of the optical and mechanical phases are sensitive to noise, the synthetic electric field, represented by the time evolution of the mechanical phase, is robust.

Even though the generation of synthetic electric fields is not prevented by quantum fluctuations, transmission (solid blue line) increases as the quantum regime is approached, see figure~\ref{trans}(c). However this increase can be explained by the larger mechanical amplitude according to the full quantum model (green solid line) than classically expected (green dashed line). For sufficiently large driving powers, the transmission, $T\approx4\tilde{B}^2/\tilde{\mathcal{E}^2}$, depends linearly on the mechanical amplitude, $\tilde{B}^2$ (see \ref{steady vdp} for the full analytical expression).

The generation of synthetic electric fields and the suppression of light propagation in the direction of higher optical frequencies is robust against noise. They are also present deep in the quantum regime where quantum fluctuations wash out the relative phase, $\varphi$, which is sensitive to noise. However, transmission in the blockaded direction is increased due to a larger mechanical amplitude, which is increased in the quantum regime compared to the classical limit.

\section{Conclusions}
We have studied synthetic dynamical gauge fields in the quantum regime. We have shown that even though the mechanical phase is very sensitive to noise and its steady-state value is washed out by quantum fluctuations, its nonlinear dynamics still results in the generation of synthetic electric fields. As a result, unidirectional light transport via synthetic dynamical gauge fields is robust against noise and it leads, also in the quantum regime, to significant isolation ratios. They are, however, reduced compared to the classical regime due to an increased mechanical amplitude. Our work paves the way for studying the quantum dynamics of synthetic gauge fields in optomechanics.

\section{Acknowledgments}
We thank O. Hart, F. Marquardt and S. Walter for useful comments. This work was supported by the European Union's Horizon 2020 research and innovation programme under grant agreement No 732894 (FET Proactive HOT).
\FloatBarrier

\appendix

\section{Synthetic electric fields for photons}\label{el field}
In this section, we summarize the key concepts of synthetic dynamical gauge fields introduced in \cite{walter2016}. We also outline how any time evolution of the mechanical phase can be interpreted as a synthetic electric field for photons as discussed in \cite{zapletal2018}.

The essential ingredient for implementing a synthetic gauge field for photons is phonon-assisted photon tunneling, $\hat{b}\hat{a}_2^{\dagger}\hat{a}_1$. During this tunneling process, photons collect the phase, $\phi$, of the mechanical oscillator, $\langle\hat{b}\rangle=Be^{i\phi}$. The gauge transformation 
\begin{align}
\phi & \rightarrow\phi + \chi_2 - \chi_1\label{trafo phi}\\
\hat{a}_j & \rightarrow\hat{a}_j e^{\chi_j},\,\,{\rm for}\,\,j=1,2,\label{trafo a}
\end{align}
yields an equivalent description of the system dynamics for any real functions $\chi_j(t)$. In this way, the mechanical phase forms a synthetic gauge field for photons. Following the gauge transformation, any constant mechanical phase can be gauged away in this rather simple setup. On the other hand, for the mechanical phase evolving in time, the gauge transformation has to be accompanied by the shift of frequencies, $\Omega \rightarrow \Omega +\dot{\chi}_1 - \dot{\chi}_2$ and $\nu_j\rightarrow\nu_j - \dot{\chi}_j$. If the mechanical oscillator performs self-sustained oscillations, i.e. $\langle\hat{b}\rangle=Be^{i\phi}e^{-i\Omega t }$ with a constant amplitude $B$, the mechanical phase $\phi$ is free to evolve according to its own dynamics. The dynamics of the mechanical phase results in the generation of a synthetic electric field, $\mathcal{E}=\dot{\phi}$, for photons \cite{zapletal2018}. 

To gain more physical intuition, consider that the mode $\hat{a}_1$ is driven by a laser. The mechanical phase can be absorbed into mode "2", $\hat{a}_2  \rightarrow \hat{a}_2e^{-i\phi}$, according to the gauge transformation (\ref{trafo phi}) and (\ref{trafo a}) by choosing $\chi_1=0$ and $\chi_2= -\phi$ and obtaining $\nu_2\rightarrow\nu_2 + \mathcal{E}$. This effective optical frequency shift represents the synthetic electric field similarly as an energy difference of electronic levels, i.e.~a voltage drop, leads to a real electric field. Note the analogy to classical electromagnetism where an electric field can be represented either by the time evolution of a vector potential or by the gradient of a scalar potential. In the same way, the synthetic electric field for photons corresponds to the time evolution of the mechanical phase or to an effective optical frequency shift.

\section{Steady states in the classical limit}\label{steady vdp}
In this section, we study the steady state in the classical limit, $\zeta\rightarrow0$, of the two optical modes coupled by phonon-assisted tunneling where the phonon mode is described by the van-der-Pol oscillator. To this end, we find the stationary solutions of the classical equations of motion (\ref{eq_bt}), (\ref{eq_akt}), and (\ref{eq_alt}). We proceed in the same way as in \cite{zapletal2018}. First, we use a time-dependent gauge transformation to represent the time evolution of the mechanical phase in a form of the effective optical frequency shift, which corresponds to the synthetic electric field. Then, we use linearity of the equations of motion for the optical modes to find their stationary values. Finally, by substituting in the stationary optical amplitudes, we obtain the stationary values of the mechanical amplitude and the effective optical frequency shift.

The evolution of the mechanical phase and the mechanical amplitude can be separated by the time-dependent gauge transformation
\begin{align}
\tilde{b} & \rightarrow  \tilde{b} e^{i\chi},\label{gauge A b}\\
\tilde{a}_1 & \rightarrow \tilde{a}_1 e^{-i\chi\,\delta_{2,k}},\label{gauge A a1}\\
\tilde{a}_2 &\rightarrow \tilde{a}_2 e^{i\chi\,\delta_{1,k}},\label{gauge A a2}
\end{align}
where $\chi=\chi(t)$ is a time-dependent gauge choice. We set $\chi=-\phi$ to express the time evolution of the mechanical phase in the form of an effective optical frequency shift,  $\tilde{\mathcal{E}}={\rm d}\chi/{\rm d}\tilde{t}$, which represents the synthetic electric field. Here, $\phi$ is the phase of the mechanical mode $\tilde{b}=\tilde{B}e^{i\phi}$. After the time-dependent gauge transformation, the mechanical phase is set to zero. Under the time-dependent gauge transformation (\ref{gauge A b},\ref{gauge A a1},\ref{gauge A a2}), the equations of motion transform to
\begin{align}
\frac{{\rm d}\tilde{B}}{{\rm d}\tilde{t}} =& \left(\delta_{1,k}-\delta_{2,k}\right)\mathcal{P}\,{\rm Im}\left [\tilde{a}^*_k \tilde{a}_l\right] +\frac{\tilde{\gamma}_1}{2}\tilde{B}-\tilde{\gamma}_2\tilde{B}^3,\label{eq_b}\\
\frac{{\rm d}\tilde{\mathcal{E}}}{{\rm d}\tilde{t}} = & \left(\tilde{\Delta}_M-\tilde{\mathcal{E}}\right)\left(1+\frac{\tilde{\gamma}_1}{2}-\tilde{\gamma}_2\tilde{B}^2\right) + \frac{\mathcal{P}}{\tilde{B}}\,{\rm Im}\left [\tilde{a}_l\right] \nonumber\\
&- \left[\left(\delta_{2,k}-\delta_{1,k}\right)\left(\tilde{\Delta}_M-2\tilde{\mathcal{E}}\right)+\tilde{\Delta}_O\right]\frac{\mathcal{P}}{\tilde{B}}\,{\rm Im}\left [\tilde{a}^*_k \tilde{a}_l\right],\label{eq_el}\\
\frac{\tilde{a}_k}{{\rm d}\tilde{t}}  = & i \tilde{\Delta}_{\rm O} \tilde{a}_k - i - i \tilde{B} \tilde{a}_{l} -\frac{1}{2}\tilde{a}_k,\label{eq_ak}\\
\frac{\tilde{a}_l}{{\rm d}\tilde{t}}  = & i \left(\delta_{2,k}-\delta_{1,k}\right)\tilde{\mathcal{E}} \tilde{a}_l - i \tilde{B} \tilde{a}_{k} -\frac{1}{2}\tilde{a}_l.\label{eq_al}
\end{align}

We study the stationary solutions of the equations of motion for $\Delta_O=\Delta_M=0$. Finite optical and mechanical detunings do not qualitatively change the steady states of the system (as it was argued in \cite{zapletal2018}). The optical amplitudes' stationary values for a given mechanical amplitude and a given synthetic electric field, read 
\begin{gather}
\tilde{a}_k=\frac{\left(\delta_{2,k}-\delta_{1,k}\right)\tilde{\mathcal{E}} + i\frac{1}{2} }{-\tilde{B}^2 - \frac{1}{4} + i \frac{1}{2}\left(\delta_{2,k}-\delta_{1,k}\right)\tilde{\mathcal{E}}},\label{ak stat}\\
\tilde{a}_l =\frac{\tilde{B}}{-\tilde{B}^2 - \frac{1}{4} + i \frac{1}{2}\left(\delta_{2,k}-\delta_{1,k}\right)\tilde{\mathcal{E}}}.\label{al stat}
\end{gather}
When the mode with the higher optical frequency,  $a_2$, is driven, the system converges to a unique steady state with vanishing synthetic electric field $\tilde{\mathcal{E}}$ for all parameters. The stationary condition for the mechanical amplitude $\tilde{B}$ reads
\begin{equation}
\left(\delta_{2,k}-\delta_{1,k}\right)\frac{\mathcal{P}}{2}+\frac{\tilde{\gamma}_1}{32}+\left( \frac{\tilde{\gamma}_1}{4}-\frac{\tilde{\gamma}_2}{16}\right)\tilde{B}^2+\left(\frac{\tilde{\gamma}_1}{2}-\frac{\tilde{\gamma}_2}{2}\right)\tilde{B}^4-\tilde{\gamma}_2\tilde{B}^6=0,\label{stat_B}
\end{equation}
which is a cubic equation for $\tilde{B}^2$ with a single real-valued root.

When the mode with the lower optical frequency, $a_1$, is driven, the synthetic electric field vanishes for small driving powers $\mathcal{P}$ and the mechanical amplitude satisfies the stationary condition (\ref{stat_B}). When driving power exceeds the threshold value
\begin{equation}
\mathcal{P}^{\rm th}=\left[ \frac{2\left(\tilde{\gamma}_1-1\right)+\tilde{\gamma}_2}{4\tilde{\gamma}_2}\right]^2,
\end{equation}
the synthetic electric field bifurcates to finite values
\begin{equation}
\tilde{\mathcal{E}}=\pm2\sqrt{\mathcal{P}-\mathcal{P}^{\rm th}}.
\end{equation}
The steady state is then bistable. Depending on the initial conditions, the synthetic electric field is either positive or negative. For the non-vanishing synthetic electric field, the stationary mechanical amplitude reads
\begin{equation}
\tilde{B}=\sqrt{\frac{\tilde{\gamma}_1-1}{2\tilde{\gamma}_2}}.\label{B above}
\end{equation}
 Above threshold, the mechanical amplitude does not depend on driving power. Importantly, light transport provided by the phonon-assisted photon tunneling is suppressed above threshold by the synthetic electric field.

To study the suppression of light propagation by the generated synthetic electric field, we use optical transmission $T$. It is the ratio of the output power leaking from the non-driven mode, $\kappa|a_2|^2$ (if mode $a_1$ is driven) or $\kappa|a_1|^2$ (if mode $a_2$ is driven), and the input power, $\kappa \mathcal{P}/\zeta^2$ \cite{zapletal2018}. The transmission
\begin{equation}
T=\frac{\tilde{B}^2}{\left(\tilde{B}^2+\frac{1}{4}\right)^2 + \frac{1}{4}\tilde{\mathcal{E}}^2}
\end{equation}
is suppressed by the synthetic electric field. As a result, the transmission in the direction to the higher optical frequency (mode $a_1$ being driven) is suppressed above threshold compared to the transmission to the lower optical frequencies (mode $a_2$ being driven).

The synthetic electric field $\tilde{\mathcal{E}}$ changes the relative phase $\varphi=\phi-\theta_2+\theta_1$ between the mechanical mode and the optical modes. The relative phase is equal to 
\begin{align}
\varphi = & \frac{\pi}{2} + \arctan\left(2\tilde{\mathcal{E}}\right),\label{phase 1}\,\,{\rm or}\\
\varphi = & -\frac{\pi}{2} + \arctan\left(2\tilde{\mathcal{E}}\right),\label{phase 1}
\end{align}
when mode $\hat{a}_1$ or mode $\hat{a}_2$, respectively, is driven.

The model of two optical modes with tunneling assisted by the van-der-Pol oscillator reduces in the classical limit to a model similar to that studied in \cite{zapletal2018}. In addition to that model, the modification of the mechanical amplitude due to radiation pressure is considered. However, this modification does not qualitatively change the generation of synthetic electric fields and unidirectional light transport.


The extent by which the mechanical amplitude is modified due to radiation pressure depends on the parameter $\tilde{\gamma}_1$, which is the ratio of the optical coherence time $1/\kappa$ and the mechanical coherence time $1/\gamma_1$. For $\tilde{\gamma}_1\ll1$, many phonons are emitted or absorbed by photons as they decay much faster than phonons. This results in a large modification of the mechanical amplitude. On the other hand, for $\tilde{\gamma}_1\gg1$, the mechanical amplitude is only marginally modified by radiation pressure since the interaction of the phonon mode with bath dominates its dynamics. However, this simultaneously suppresses phonon-assisted photon tunneling. To study the dynamics of the synthetic gauge field represented by the mechanical phase, we want to eliminate the modification of the mechanical amplitude but, at the same time, to preserve phonon-assisted photon tunneling. To this end, we set $\tilde{\gamma}_1=10$ in the main text, which provides a good trade-off between eliminating the modification of the mechanical amplitude and preserving phonon-assisted photon tunneling.

\section{Propagation to lower optical frequencies}\label{higher freq}
\begin{figure}[t]
\centering
\includegraphics[width=0.9\linewidth]{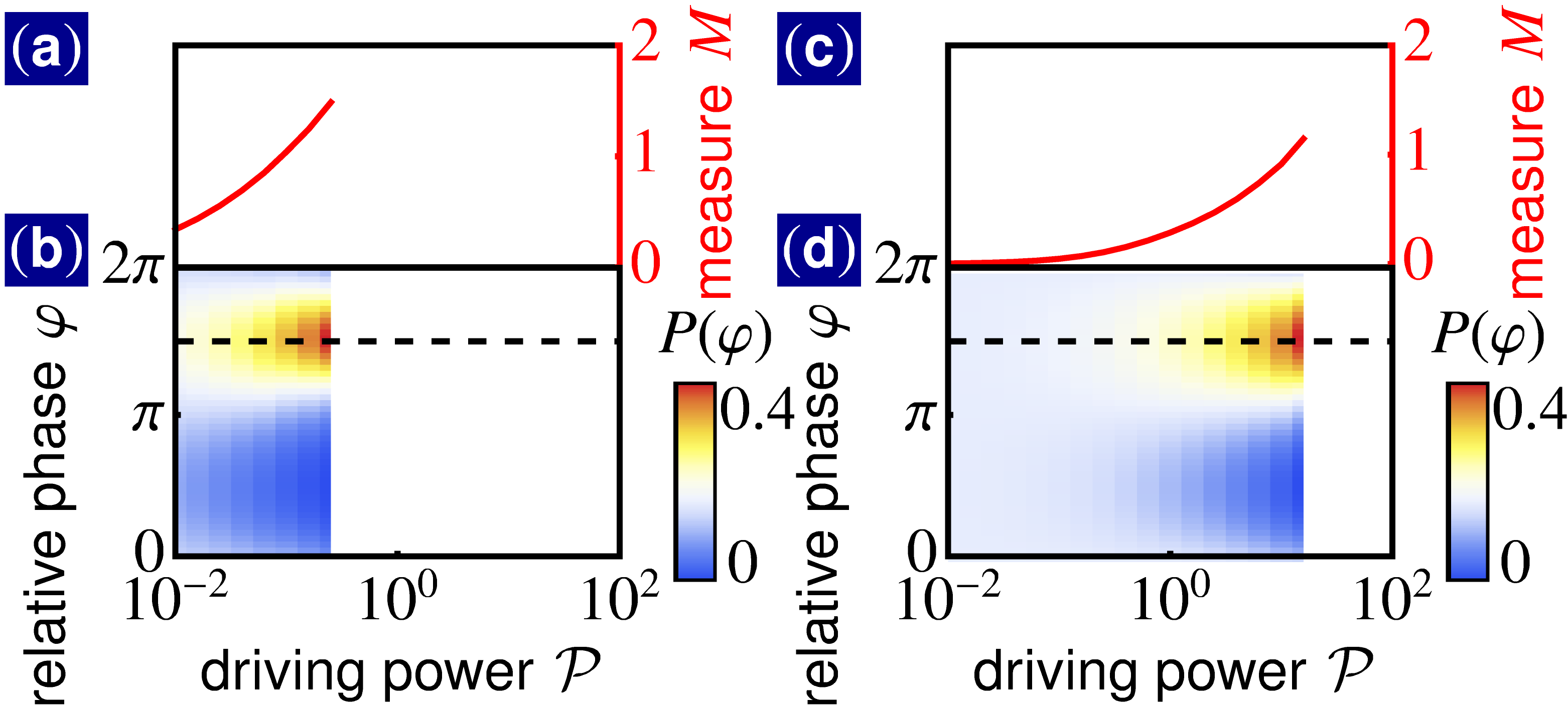}
\caption{Relative phase between mechanical and optical modes in the case when the mode, $\hat{a}_2$, with the higher optical frequency is driven. Relative phase distribution as a function of driving power $\mathcal{P}$: (b) close to the classical limit, and (d) in the quantum regime. Corresponding phase coherence measure: (a) close to the classical limit and (c) in the quantum regime. The black dashed lines show the classical value of the relative phase. (Master equation simulations for: (a) and (b) $\tilde{\gamma}_1=10$, $\tilde{\gamma}_2=200$, and $\zeta = 0.1$; and (c) and (d) $\tilde{\gamma}_1=10$, $\tilde{\gamma}_2=40$, and $\zeta = 1$.)}
\label{Pphi higher}
\end{figure}

\begin{figure}[t]
\centering
\includegraphics[width=0.9\linewidth]{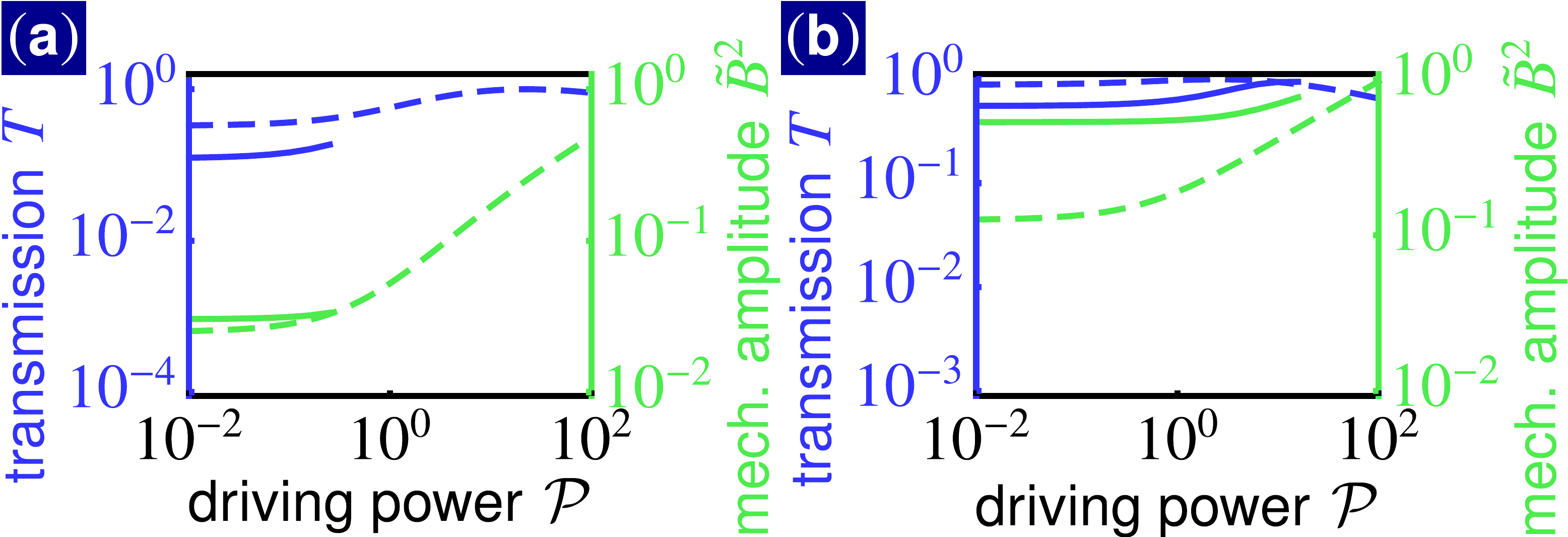}
\caption{Light propagation through the device in the case when the mode, $\hat{a}_2$, with the higher optical frequency is driven. Optical transmission $T$ and expectation value of the mechanical amplitude $\tilde{B}^2=\zeta^2\langle\hat{b}^{\dagger}\hat{b}\rangle$ as a function of driving power $\mathcal{P}$ at the onset of a finite synthetic electric field:  (a) close to the classical limit and (b) in the quantum regime. The dashed lines show classical limit. (Master equation simulations for: (a) and (b) $\tilde{\gamma}_1=10$, $\tilde{\gamma}_2=200$, and $\zeta = 0.1$;  and (c) and (d) $\tilde{\gamma}_1=10$, $\tilde{\gamma}_2=40$, and $\zeta = 1$.)}
\label{trans higher}
\end{figure}

In this section, we analyze propagation to lower optical frequencies where no synthetic electric fields are generated. We assume the model analyzed in the main text, now in the case when the higher-frequency optical mode is driven.

Results of the analysis are shown in figure~\ref{Pphi higher} and figure~\ref{trans higher}. One can see in figure~\ref{Pphi higher}(b) and figure~\ref{Pphi higher}(d) that the synthetic electric field always vanishes as the relative phase distribution is peaked around the value, $\varphi=3\pi/2$, for all driving powers. This happens both close to the classical limit, $\zeta=0.1$, and deep in the quantum regime, $\zeta=1$. With increasing driving power the relative phase distribution is more peaked around the value, $\varphi=3\pi/2$, and consequently the phase coherence measure increases, see figure~\ref{Pphi higher}.

Close to the classical limit, numerical simulations are restricted to small driving powers since many energy levels of the van-der-Pol oscillator have to be taken into account. The computational demand is even larger than in the case if the lower-frequency optical mode  is driven. In contrast to that case, many energy levels of the non-driven optical mode are populated as the phonon-assisted photon tunneling is not suppressed for the vanishing synthetic electric field, see figure~\ref{trans higher}.

For $\zeta=0.1$, when the quantum fluctuations are moderately strong, the mechanical amplitude according to the full quantum model (solid green line) is well approximated by its classical limit as it can be seen in figure~\ref{trans higher}(a). However, in the quantum regime, for $\zeta=1$, only a few mechanical energy levels are populated and their quantization has to be taken into account. Since the quantization of energy levels is neglected for the classical equations of motion, they predict significantly different mechanical amplitude than the full quantum model, see figure~\ref{trans higher}(b).


The transmission to the lower-frequency optical mode is decreased by quantum fluctuations for any value of driving power both close to the classical limit and deep  in the quantum regime, see figure~\ref{trans higher}. The transmission slightly increases with increasing driving power. This is in contrast to the large suppression of transmission to higher optical frequencies caused by the synthetic electric field (figure~\ref{trans}). As a result, significant isolation ratios of transmission to the lower and to the higher optical frequencies can be achieved.

\section{Numerical master equation simulations}\label{simulations}
In this section, we provide details about master equations simulations, with which the main result of this manuscript were obtained.

The starting point of our analysis is the master equation (\ref{master}). This equation can be rewritten as
\begin{equation}
\dot{\hat{\rho}} = \mathcal{L}\hat{\rho},
\end{equation}
in terms of the superoperator $\mathcal{L}$. We solve this equation for a steady state satisfying the equation
\begin{equation}
\mathcal{L}\hat{\rho}=0.\label{master SS}
\end{equation}
The system described by equation (\ref{master}) has a unique steady state.

To numerically solve equation (\ref{master SS}), we express the superoperator $\mathcal{L}$ in the truncated Fock basis taking into account $N_1$, $N_2$ and $N_m$ lowest Fock states of modes $\hat{a}_1$, $\hat{a}_2$ and $\hat{b}$, respectively. The dimensions of the matrix $\mathcal{L}$, $\left(N_1\,N_2\,N_m\right)^2 \times \left(N_1\,N_2\,N_m\right)^2$, rapidly increase with the size of the truncated Hilbert space. This represents the major restriction on occupations of optical and mechanical modes. To partially bypass this restriction, we make use of the displacement transformation
\begin{equation}
\hat{\rho} \rightarrow \mathcal{D}^{\dagger}(\alpha_1,\alpha_2)\hat{\rho} \mathcal{D}(\alpha_1,\alpha_2),
\end{equation}
where 
\begin{equation}
\mathcal{D}(\alpha_1,\alpha_2)=e^{\alpha_1 \hat{a}_1^{\dagger} - \alpha_1^* \hat{a}_1}\,e^{\alpha_2 \hat{a}_2^{\dagger} - \alpha_2^* \hat{a}_2},
\end{equation}
is the displacement operator and $\alpha_j$ are complex numbers. The displacement transformation allows numerical simulations of strongly driven optical modes in the case when they are in a state close to a coherent state.

The master equation simulations were efficient in the case of the lower optical frequency being driven. The dynamically generated synthetic electric field suppresses the transmission to the higher-frequency optical mode. As a result, the occupation of the higher-frequency optical mode remains small and the lower-optical frequency mode is in a state close to a coherent state. In this case, the displacement transformation allows simulating a very strong coherent driving. In the case of the higher frequency being driven, the transmission is not suppressed, which leads to a large occupations also of the lower-frequency optical mode. As a result, master equation simulations are restricted to lower driving powers especially close to the classical regime where the occupation of the mechanical self-oscillator is large.

The figures in this manuscript were produced for the following truncations: figures \ref{Pphi}(a), \ref{Pphi}(b) and \ref{trans}(a) for $\left(N_1=3,N_2=15,N_m=15\right)$; figures \ref{Pphi}(c), \ref{Pphi}(d) and \ref{trans}b for $\left(N_1=5,N_2=5,N_m=5\right)$; figures \ref{Pphi}(e), \ref{Pphi}(f) and \ref{trans}(c) for $\left(N_1=3,N_2=15,N_m=15\right)$; figures \ref{Pphi higher}(a), \ref{Pphi higher}(b) and \ref{trans higher}(a) for $\left(N_1=18,N_2=4,N_m=11\right)$; \ref{Pphi higher}(c), \ref{Pphi higher}(d) and \ref{trans higher}(b) for $\left(N_1=30,N_2=7,N_m=5\right)$ as well as figure \ref{spectrum} for $\left(N_1=5,N_2=5,N_m=5\right)$.

\section{Optical power spectrum in the quantum regime}\label{app spectrum}
In this section, we study the power spectrum of the non-driven optical mode in the quantum regime. The shift in frequency of this mode due to the time evolution of the mechanical phase corresponds the synthetic electric field \cite{zapletal2018}.

\begin{figure}[t]
\centering
\includegraphics[width=0.45\linewidth]{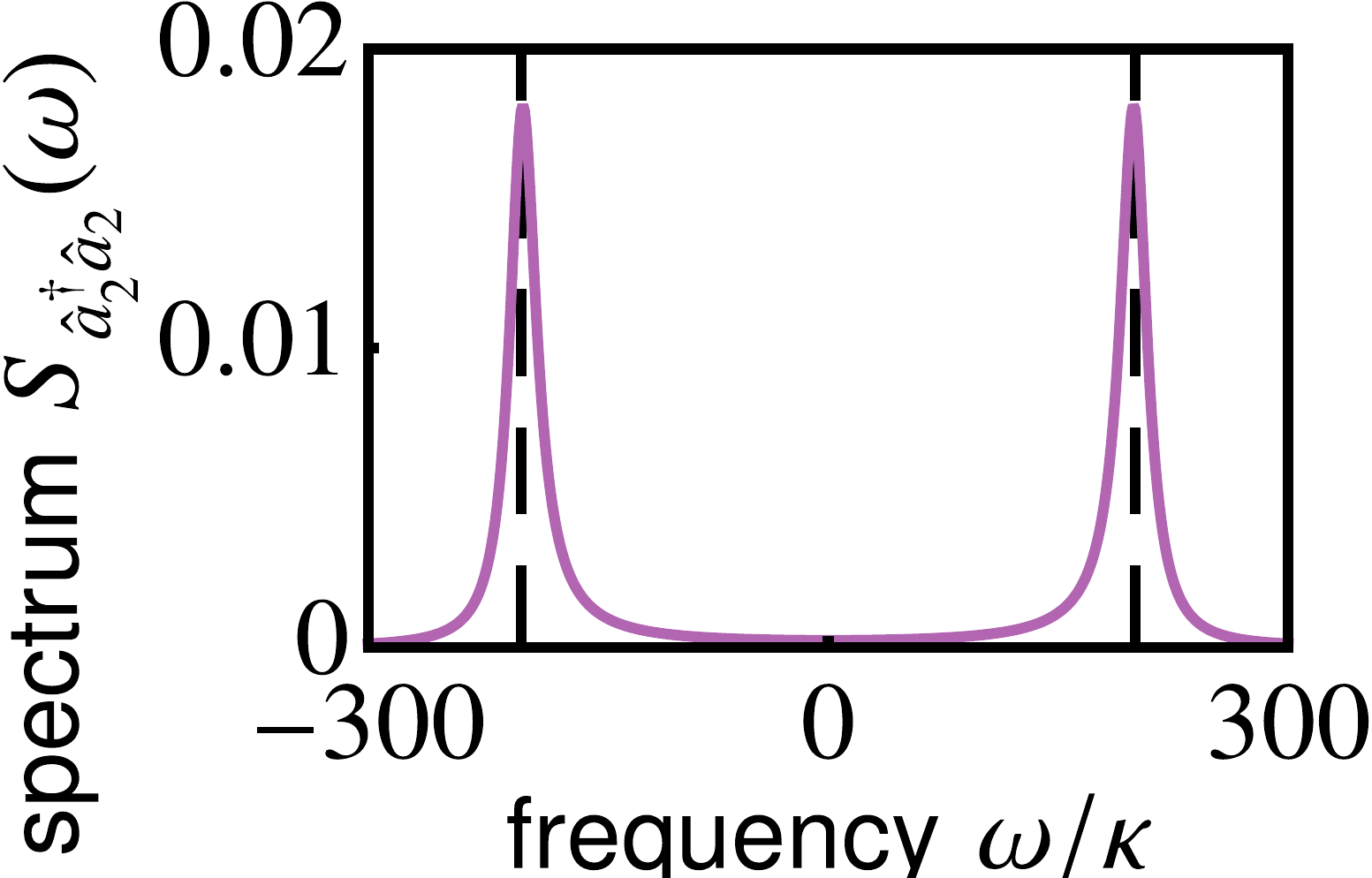}
\caption{Power spectrum, $S_{\hat{a}_2^{\dagger}\hat{a}_2}\left( \omega \right)$, of the non-driven optical mode, $\hat{a}_2$, in the frame rotating with the resonant frequency, $\nu_2$. The black dashed lines show the values of the synthetic electric field $\tilde{\mathcal{E}}$ according to the classical equations of motion. (Master equation simulations for:  $\tilde{\gamma}_1=10$, $\tilde{\gamma}_2=200$, $\zeta = 1$, and $\mathcal{P}=10000$.)}
\label{spectrum}
\end{figure}

The power spectrum of the optical mode, $\hat{a}_2$, reads
\begin{equation}
S_{\hat{a}_2^{\dagger}\hat{a}_2}\left( \omega \right)=\int{\rm d}t\,e^{i\omega t}\langle \hat{a}_2^{\dagger}(t)\hat{a}_2(0)\rangle,
\end{equation}
where $\langle \hat{a}_2^{\dagger}(t)\hat{a}_2(0)\rangle={\rm Tr}\left[ \hat{a}_2^{\dagger}(t)\hat{a}_2(0) \hat{\rho}_{SS}\right]$ and $\hat{\rho}_{SS}$ is the steady state density matrix. The power spectrum $S_{\hat{a}_2^{\dagger}\hat{a}_2}\left( \omega \right)$, can be efficiently computed in the quantum regime where only few optical and mechanical energy levels are occupied. The spectrum is depicted in figure~\ref{spectrum} for the case where mode, $\hat{a}_1$, is driven by a laser of large power. In this case, a significant suppression of light transmission is reached, as discussed in the main text. One can see that the spectrum features two peaks at frequencies shifted from the resonant frequency ($\omega=0$ in the rotating frame). The peaks are located at frequencies corresponding to the classical values of the synthetic electric field $\tilde{\mathcal{E}}\approx \pm 2 \sqrt{\mathcal{P}}$. This shows that the synthetic electric field is generated also in the quantum regime and its values agree with the ones predicted by classical equations of motion.

The power spectrum of the non-driven optical mode demonstrates that the synthetic electric field is robust against noise. This is in contrast to the relative phase, $\varphi$, which is washed out by quantum fluctuations in the quantum regime. 

\section*{References}
\bibliographystyle{iopart-num}
\bibliography{Paper_vdp}

\end{document}